\documentclass[11pt]{article}
\usepackage{amssymb}
\usepackage{amsfonts}

\usepackage{graphics,epsfig}
\usepackage{graphicx}
\usepackage{latexsym}
\usepackage{amsmath, amsthm, amsfonts, amssymb}
\usepackage{verbatim}

\newtheorem{theorem}{Theorem}

\newtheorem{lemma}[theorem]{Lemma}

\begin{document}

\title{An upper bound for the KS--entropy in quantum mixing systems}

\author{Ignacio S. Gomez}
\date{}
\maketitle

\begin{small}
\begin{center}
IFLP, UNLP, CONICET, Facultad de Ciencias Exactas, Calle 115 y 49, 1900 La Plata, Argentina\\
\end{center}
\end{small}
\begin{abstract}
We present an upper bound for the Kolmogorov--Sinai entropy of quantum systems having a mixing quantum phase space. The method for this estimation is based on the following ingredients: i) the graininess of quantum phase space in virtue of the Uncertainty Principle, ii) a time rescaled KS--entropy that introduces the characteristic timescale as a parameter, and iii) the factorization property of the mixing correlations.
The analogy between the structures of the mixing level of the ergodic hierarchy and of its quantum counterpart is shown. Moreover, the logarithmic timescale, characteristic of quantum chaotic systems, is obtained.
\end{abstract}
\noindent

\begin{small}
\noindent \centerline{Keywords: \ \ KS--entropy \ -- \ Mixing \ -- \ Graininess \ -- \ Logarithmic timescale}
\end{small}

\bibliography{pom}

\section{Introduction}
\label{intro}

Kolmogorov--Sinai entropy (KS--entropy) is considered as one of the most robust and significant indicator of chaos, theoretically and for applications \cite{Tab79,Wal82,Gutz90,Lic92}.
Basically, the KS-entropy
assigns measures to bunches of trajectories and computes
the Shannon-entropy per time-step of the ensemble of
bunches in the limit of infinitely many time-steps. Moreover,
the Pesin theorem \cite{Pes77,You03} links the KS-entropy with the
Lyapunov coefficients which are a measure of the exponential instability, i.e. they characterize the chaotic motion.
In turn, in classical mechanics the two properties necessary for
chaos to occur are a continuous spectrum and a continuous
phase space \cite{Cas95}.

However, in quantum mechanics the arising of chaos is more subtle. Firstly,
the most of quantum systems which present chaotic features in its classical limit
have discrete spectrum and secondly, the Correspondence
Principle CP implies the transition from quantum to classical
mechanics for all phenomena including the chaos. Furthermore, by the Uncertainty Principle the quantum phase space is discrete and divided into elementary cell of finite size, which constitutes the so called \emph{graininess}.
Despite these difficulties, the KS--entropy allows one to give a concrete answer about the emergence of chaos in the classical limit, i.e. the quantum chaos \cite{Ber89,Sto99,Haa01}. The key point is that one can model the behavior of classical
chaotic systems of continuous spectrum from classical discretized
models in such way that the KS--entropies of the continuous and the discrete
one tend to coincide for a
certain appropriate range \cite{Cri93,Slo94,Cri94,Fal03,Ben04,Ben05}. This remark is crucial in order to obtain the characteristic timescales of quantum chaos
where the classically behavior and the chaotic one overlap each other \cite{Hel89}.

On the other hand, many chaotic systems of interest are mixing, i.e. the subsets of phase space have a correlation decay such that any two subsets are statistically independent for large times \cite{Wal82,Las85,Bel97,Ber06}. This property is one of the most useful concepts to describe phenomena such as chaos, approach to equilibrium and relaxation in dynamical systems theory \cite{Cas95}. In a series of works quantum extensions of the mixing property were proposed \cite{paper0,paper1,paper2,paper3,paper4,paper5}, from which we characterized
the chaotic behaviors of the Casati--Prosen model \cite{paper1,Cas05} and the kicked rotator \cite{Sto99,Haa01,paper1}, and recently the Gaussian Orthogonal Ensembles were obtained \cite{paper5}.

The main goal of this paper is to obtain an expression of the KS--entropy for quantum systems having a mixing classical analogue, making use of the quantum phase space graininess and the mixing property. Moreover, our approach allows one to shed light on the foundations of the quantum chaos. In particular, we obtain the logarithmic timescale as a consequence of the formalism presented.

The paper is organized as follows. In Section \ref{sec prelim} we give the preliminaries that
we employ throughout the paper. In Section \ref{sec mixing} we prove some properties of the mixing correlations which are the key to obtain the KS--entropy. In Section \ref{sec KS quantum mixing} we give an upper bound of the KS--entropy obtained by means of the discretized quantum phase space and the time--rescaling property of the KS--entropy. Here we obtain the logarithmic timescale as a consequence of the estimation of the KS-entropy.
Finally, in Section \ref{sec conclusions} we discuss and draw some conclusions.

\section{Preliminaries}\label{sec prelim}

We give the notions and concepts to develop the results of the paper. First of all, we clarify the notation we will use throughout the paper.

We denote by $\langle\hat{O}\rangle_{\hat{\rho}}$ the mean value of observable $\hat{O}$ when the system is in state $\hat{\rho}$, i.e. $\langle\hat{O}\rangle_{\hat{\rho}}=Tr(\hat{\rho} \hat{O})$ where $Tr(  \ . \ )$ is the trace operator. If $\hat{\rho}$ is any initial state at time $t=0$, we denote by $\hat{\rho}(t)$ the state at time $t$, i.e. $\hat{\rho}(t)=\hat{U}\hat{\rho}\hat{U}^{\dag}$ where $\hat{U}=e^{-i\frac{\hat{H}t}{\hbar}}$ is the well-known operator evolution for Hamiltonian $\hat{H}$, and $\hat{U}^{\dag}$ is its adjoint operator.

\subsection{Kolmogorov--Sinai entropy}\label{subsec KS}

We recall the definition of the KS--entropy within the standard framework of measure theory \cite{Wal82,Las85,Ber06}.
Consider a dynamical system given by $(\Gamma, \Sigma, \mu, \{T_t\}_{t\in J})$, where $\Gamma$ is the phase space, $\Sigma$ is a $\sigma$-algebra, $\mu:\Sigma \rightarrow [0,1]$ is a normalized measure and $\{T_t\}_{t\in J}$ is a semigroup of preserving measure transformations. For instance, $T_t$ could be the classical Liouville transformation or the corresponding classical transformation associated to the quantum Schr\"{o}dinger transformation. $J$ is usually $\mathbf{R}$ for continuous dynamical systems and $\mathbf{Z}$ for discrete ones.

Let us divide the phase space $\Gamma$ in a partition $Q$ of $m$ small cells $A_{i}$ of measure $\mu (A_{i})$. The entropy of $Q$ is defined as
\begin{equation}\label{entropy partition}
H(Q)=-\sum_{i=1}^{m}\mu(A_{i})\log\mu(A_{i}).
\end{equation}
Now, given two partitions $Q_1$ and $Q_2$ we can obtain the partition $Q_1\vee Q_2$ which is $\{a_i\cup b_j: a_i\in Q_1, b_j\in Q_2\}$, i.e. $Q_1\vee Q_2$ is a refinement of $Q_1$ and $Q_2$.
In particular, from $Q$ we can obtain the partition $H(\vee_{j=0}^{n}T^{-j}Q)$ being $T^{-j}$ the inverse of $T_{j}$ (i.e. $T^{-j}=T_{j}^{-1}$) and $T^{-j}Q=\{T^{-j}a:a\in Q\}$.
From this, the KS--entropy $h_{KS}$ of the dynamical system is defined as
\begin{equation}\label{KS-entropy}
h_{KS}=\sup_{Q}\{\lim_{n\rightarrow\infty}\frac{1}{n}H(\vee_{j=0}^{n}T^{-j}Q)\}
\end{equation}
where the supreme is taken over all measurable initial partitions $Q$ of $\Gamma$.
From the viewpoint of information theory, the Brudno theorem says that the KS--entropy is the average unpredictability of information of all possible trajectories in the phase space. Furthermore, Pesin theorem relates the KS--entropy with the exponential instability of motion given by the Lyapunov exponents. Then, the main content of Pesin theorem is that $h_{KS}>0$ is a sufficient condition for chaotic motion.
\subsection{Time rescaled KS--entropy}
By taking $(\Gamma, \Sigma, \mu, \{T_t\}_{t\in J})$ as the classical analogue of a quantum system and considering the timescale $\tau$ within the quantum and classical descriptions coincide \cite{Cas95,Hel89}, the definition (\ref{KS-entropy}) can be expressed as
\begin{equation}\label{KS-entropy1}
h_{KS}=\sup_{Q}\{\lim_{n\tau\rightarrow\infty}\frac{1}{n\tau}H(\vee_{j=0}^{n\tau}T^{-j}Q)\}
\end{equation}
Now since $T^{-j\tau}=(T_{\tau})^{-j}$ one can recast (\ref{KS-entropy1}) as
\begin{equation}\label{KS-entropy2}
h_{KS}=\frac{1}{\tau}\sup_{Q}\{\lim_{n\rightarrow\infty}\frac{1}{n}H(\vee_{j=0}^{n}(T_{\tau})^{-j}Q)\} \nonumber
\end{equation}
Finally, from this equation one can express $h_{KS}$ as
\begin{equation}\label{KS-entropy3}
h_{KS}=\frac{1}{\tau}h_{KS}^{(\tau)}   \ \ \ , \ \ \ h_{KS}^{(\tau)}=\sup_{Q}\{\lim_{n\rightarrow\infty}\frac{1}{n}H(\vee_{j=0}^{n}(T_{\tau})^{-j}Q)\}
\end{equation}
The main role of the time rescaled KS--entropy $h_{KS}^{(\tau)}$ is that allows to introduce the timescale $\tau$ as a parameter. This concept will be an important ingredient for obtaining the logarithmic timescale.

\subsection{Weyl--Wigner--Moyal formalism}\label{subsec WWW}
\label{sec:2}
We review some properties of the Weyl symbol and the Wigner function for the development of next sections \cite{Bay78,Wig84,Dit02}.
If $\hat{A}$ is an operator then the Weyl symbol of $\hat{A}$ is a distribution function over phase space defined by \cite{Wig84,Dit02}
\begin{eqnarray}\label{wigner1}
\widetilde{W}_{\hat{A}}(q,p)=\int_{\mathbf{R}}\langle q+\frac{\Delta}{2}|\,\hat{A}
\,|q-\frac{\Delta}{2}\rangle e^{-i\frac{p\Delta}{\hbar}}d\Delta
\end{eqnarray}
The Wigner function of $\hat{A}$ is defined by means of its Weyl symbol as
\begin{eqnarray}\label{wigner1bis}
W_{\hat{A}}(q,p)=\frac{1}{h}\widetilde{W}_{\hat{A}}(q,p)=\frac{1}{h}\int_{\mathbf{R}}\langle q+\frac{\Delta}{2}|\,\hat{A}\,|q-\frac{\Delta}{2}\rangle e^{-i\frac{p\Delta}{\hbar}}d\Delta
\end{eqnarray}
where $h$ is the Planck constant. The Wigner function has a relevant property that allows one to express any quantum mean value as an integral in phase space \cite{Wig84}
\begin{eqnarray}\label{wigner8}
\langle\hat{O}\rangle_{\hat{\rho}}=
\int_{\mathbf{R}} dqdp \ W_{\hat{\rho}}(q,p)\widetilde{W}_{\hat{O}}(q,p)
\end{eqnarray}
\section{Mixing correlations}\label{sec mixing}
We present some results about the mixing correlations, classical and quantum, that we will use in the next sections.
\subsection{Classical correlations}\label{subsec classical mixing}
In ergodic theory \cite{Wal82,Las85,Ber06}, correlation decay of mixing systems is the most important property for the validity of the statistical description
because different regions of phase space become statistical independent when they are enough separated in time. More precisely, if we have a dynamical system $(\Gamma,\mu,\Sigma,\{T_t\})$ where $\Gamma$ is the phase space, $\mu:\Sigma \rightarrow [0,1]$ is a normalized measure and $\{T_t\}_{t\in J}$ is a semigroup of preserving measure transformations then the mixing correlations are mathematically expressed as
\begin{equation}\label{mixing1}
lim_{t\rightarrow\infty}C(T_tA,B)=lim_{t\rightarrow\infty}\mu(T_tA \cap B)-\mu(A)\mu(B)=0
\end{equation}
for all $A,B\in\Gamma$. The eq. (\ref{mixing1}) expresses the so called \emph{mixing property} which is satisfied by several examples like Sinai billiards, Brownian motion, chaotic maps, etc \cite{Cas95,Sto99,Haa01,Cas05}.

The Frobenius--Perron operator $P_t:L^1(\Gamma)\rightarrow L^1(\Gamma)$ associated to the transformation $T_t:\Gamma\rightarrow\Gamma$ is given by \cite{Las85}
\begin{eqnarray}\label{1-1bis}
\int_{A}P_t\phi=\int_{T_t^{-1}(A)}\phi \nonumber
\end{eqnarray}
for all $\phi\in L^1(\Gamma)$ and $A\subseteq\Gamma$, where $T_t^{-1}(A)$ is the preimage of $A$.
Any normalized distribution $f_*\in L^1(\Gamma)$ such that $P_tf_*=f_*$
is called a \emph{fixed point} of $P_t$. Furthermore, it can be shown that $f_*$ is a fixed point of $P_t$ if and only if the measure $\mu_*(A)=\int_{A}f_*$ is invariant under $T_t$ \cite{Las85}, i.e. $\mu_*(T_t A)=\mu_*(A)$. For this reason the distribution $f_*$ is also frequently called \emph{invariant density}. From now on we use both names indistinctly to refer us to $f_*$.

Assuming that the Frobenius-Perron operator $P_t$ associated with each transformation
$T_t$ has a fixed point $f_{\ast}$ then the following relevant property of mixing systems can be deduced. In the following we present some results whose proofs can be found in the appendix.
\begin{lemma}\label{lemma1}
Let $f_{\ast}$ be a normalized distribution which is a fixed point of the Frobenius-Perron operator $P_t$
and let $1_{A_1}, 1_{A_2},...,1_{A_n}:\Gamma\rightarrow \mathbf{R}$ be characteristic functions. Then we have
\begin{eqnarray}\label{lemma1-4}
\int_\Gamma f_{*} 1_{A_1}\cdots1_{A_n}=\left(\int_\Gamma f_{*} 1_{A_1} \right)\cdots\left(\int_\Gamma f_{*} 1_{A_n} \right)
\end{eqnarray}
\end{lemma}
This lemma expresses that the classical mean value of a product can be factorized in the corresponding product of each mean value where the probability density $f_{\ast}$ is a fixed point of $P_t$. The ``factorization property" of eq. (\ref{lemma1-4}) will be useful to obtain the KS--entropy expressed in terms of mean values, but first we must explore its consequences in the context of quantum mixing correlations. We will see below how to do this.
\subsection{Quantum correlations}\label{subsec quantum mixing}
A quantum counterpart of the mixing correlation of (\ref{mixing1}) was derived in \cite{paper0,paper1}, with a decay correlation between states and observables rather than between subsets of phase space, given by
\begin{eqnarray}\label{qmixing1}
\lim_{t\rightarrow\infty}C(\hat{\rho}(t),\hat{O})=\lim_{t\rightarrow\infty}\left(\langle\hat{O}\rangle_{\hat{\rho}(t)}-\langle\hat{O}\rangle_{\hat{\rho}_{*}}\right)=0
\end{eqnarray}
where the role played by the subsets $A,B$ now is played by the states and the observables $\hat{\rho}(t)$, $\hat{O}$.
The eq. (\ref{qmixing1}) describes the relaxation of any initial quantum state $\hat{\rho}$ with a weak limit $\hat{\rho}_{\ast}$ where the relaxation is understood in the sense of the mean values, i.e. the decoherence of observables \cite{VKam54,Van55,Dan62,Zeh73,Omn94,Omn99}. Moreover, we can show that
the steady state $\hat{\rho}_{*}$ is the quantum analogue of the invariant density $f_{\ast}$ of the lemma \ref{lemma1}, which is the content of the following lemma.
\begin{lemma}\label{lemma2}
The state $\hat{\rho}_{\ast}$ is a fixed point of the evolution operator $\hat{U}_t=e^{-it\frac{\hat{H}}{\hbar}}$ being $\hat{H}$ the Hamiltonian of the quantum system, i.e. $\hat{U}_t\hat{\rho}_{\ast}\hat{U}_t^{\dag}=\hat{\rho}_{\ast}$.
\end{lemma}
\noindent From the Lemma \ref{lemma2} one can prove its analogue version in phase space.
\begin{lemma}\label{lemma3}
The Wigner distribution $W_{\hat{\rho}_{\ast}}(q,p)$ of $\hat{\rho}_{\ast}$ is a fixed point of
the Frobenius-Perron operator $P_t$ associated with the classical evolution $T_t$ given by Hamiltonian equations.
\end{lemma}
\begin{figure}[th]
\begin{center}
\includegraphics[width=12cm]{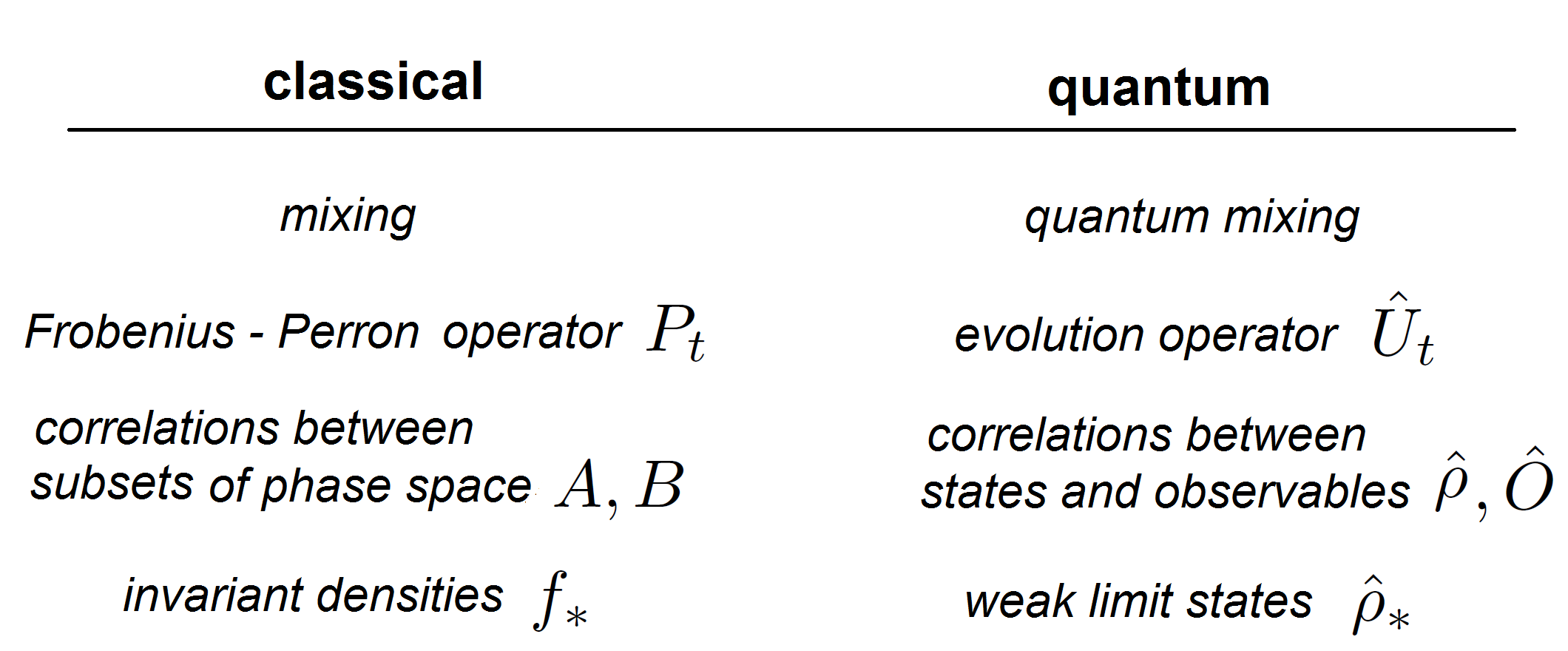}
\end{center}
\caption{An scheme showing the structures of the mixing level and of its quantum counterpart.}
\label{fig:classical-quantum}\end{figure}
In Fig. \ref{fig:classical-quantum} it can be seen the similarities between the classical and quantum structures of the mixing level of the ergodic hierarchy. Each classical concept has its associated quantum analogue and therefore, the analogy is total.

\section{KS--entropy in the context of quantum mixing systems}\label{sec KS quantum mixing}
Having established some properties of the mixing correlations and taking into account the graininess of quantum phase space, now we are able to
give an expression of the KS--entropy. We begin by employing the mixing correlations described in Section \ref{subsec classical mixing}.

For the sake of simplicity we consider a bidimensional\footnote{It should be noted that the results can be generalized to any dimension of phase space.} and discretized quantum phase composed by rigid cells of minimal size $\Delta q\Delta p=h$ with $h$ the Planck constant. We assume that the dynamics in phase space is mixing and therefore, chaotic. In particular, this implies that the systems occupies a bounded compact region $\Omega\in\mathbf{R}^2$ with $\mu(\Omega)<\infty$. We also can consider that $\mu(\Omega)=1$ (otherwise $\tilde{\mu}=\frac{\mu}{\mu(\Omega)}$ is normalized).

\begin{figure}[th]
\begin{center}
\includegraphics[width=8cm]{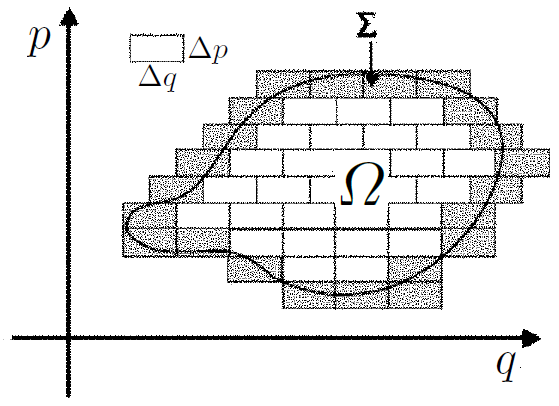}
\end{center}
\caption{Bounded motion and graininess in quantum phase space. In the semiclassical limit $q\gg1$ the region $\Omega$ that the system occupies has a volume that is approximately the sum of the volumes of the rigid boxes $\Delta q\Delta p$ contained in $\Omega$. The region $\Sigma$ corresponding to the rigid boxes that intersect the frontier of $\Omega$ can be neglected in the limit $q\gg1$.}
\label{fig:grain}\end{figure}

By the Uncertainty Principle it follows that there exists a maximal partition\footnote{That is, the greatest refinement that one can take.} $Q_{max}=\{A_1,\ldots,A_M\}$ of $\Omega$ composed by $M$ identical rectangle cells $A_i$ of dimensions $\Delta q\Delta p$ and $\mu(A_i)=h$ for all $i=1,\ldots,M$ where $M$ is the maximal number of cells $A_i$
that intersect $\Omega$. An illustration is shown in Fig. \ref{fig:grain} where $q=\frac{\mu(\Omega)}{h}\geq1$ is the well known quasiclassical parameter that ``measures" how far or near is the quantum system of its classical limit. In this sense, the relation $q\gg1$ characterized the semiclassical limit.
Since $\mu(\Omega)=1$ and $Q_{max}$ is a partition we have that $\sum_{i=1}^M \mu(A_i)=\sum_{i=1}^M h=1$, that is
\begin{eqnarray}\label{graininess}
Mh=1
\end{eqnarray}
Eq. (\ref{graininess}) expresses the graininess of the quantum phase space.

In order to obtain $h_{KS}$ the key point is to calculate $h_{KS}^{(\tau)}$ where $\tau$ is the timescale in which the classical and quantum descriptions overlap. For accomplish this, one has to consider $T_{\tau}$ instead of $T$. Thus, the supreme 
in (\ref{KS-entropy2}) can be replaced by
$\lim_{n\rightarrow\infty}\frac{1}{n}H(\vee_{j=0}^{n}T_{\tau}^{-j}Q_{max})$ in the context of the graininess of the quantum phase space. Now,
the partition $\vee_{j=0}^{n}T_{\tau}^{-j}Q_{max}$
is given by
\begin{eqnarray}\label{4-1}
&\vee_{j=0}^{n}T_{\tau}^{-j}Q_{max}=\nonumber\\
&\{A_{i_0}\cap T_{\tau}^{-1}A_{i_1} \cap \ldots \cap T_{\tau}^{-n}A_{i_n}:i_l=1,\ldots,M \ ; \ l=1,\ldots,n\}
\end{eqnarray}
Given a $(n+1)$--upla $(i_0,i_1,\ldots,i_n)\in \{1,\ldots,M\}^{(n+1)}$ and since the dynamics is bounded and contained in the compact $\Omega$ then one has that $T_{\tau}^{-l}A_{i_l}\subset \Omega$ for all $l=0,\ldots,n$. Thus, one can express $\mu(A_{i_0}\cap T_{\tau}^{-1}A_{i_1} \cap \ldots \cap T_{\tau}^{-n}A_{i_n})$ as
\begin{eqnarray}\label{4-1}
&\mu(A_{i_0}\cap T_{\tau}^{-1}A_{i_1} \cap \ldots \cap T_{\tau}^{-n}A_{i_n})=\nonumber\\
&\int_{\Omega}1_{\Omega}(q,p)1_{A_{i_0}}(q,p)1_{T_{\tau}^{-1}A_{i_1}}(q,p)\cdots1_{T_{\tau}^{-n}A_{i_n}}(q,p)dqdp
\end{eqnarray}
Since $\mu(\Omega)=1$ it is clear that $1_{\Omega}(q,p)$ is a normalized distribution. Moreover,
$1_{\Omega}(q,p)$ is a fixed point of the Frobenius--Perron operator $P_t$ associated with the transformation $T_t$ due to the measure
$\mu_{\Omega}(A)=\int_{A}1_{\Omega}$ is trivially $\mu$ which by definition is invariant under $T_t$.

Then, given the distribution $1_{\Omega}(q,p)$ and the characteristic functions \\
$1_{A_{i_0}}(q,p),1_{T_{\tau}^{-1}A_{i_1}}(q,p),\ldots,1_{T_{\tau}^{-n}A_{i_n}}(q,p)$  one can apply the Lemma \ref{lemma1}, thus obtaining
\begin{eqnarray}\label{4-3}
&\int_{\mathbf{R}^2} 1_{\Omega}1_{A_{i_0}}1_{T_{\tau}^{-1}A_{i_1}}\cdots1_{T_{\tau}^{-n}A_{i_n}}dqdp=\nonumber\\
&\left(\int_{\mathbf{R}^2} 1_{\Omega}1_{A_{i_0}}dqdp\right)\left(\int_{\mathbf{R}^2} 1_{\Omega}1_{T_{\tau}^{-1}A_{i_1}}dqdp\right)\cdots\left(\int_{\mathbf{R}^2} 1_{\Omega}1_{T_{\tau}^{-n}A_{i_n}}dqdp\right)
\end{eqnarray}
That is,
\begin{eqnarray}\label{4-3}
&\int_{\mathbf{R}^2} 1_{A_{i_0}}1_{T_{\tau}^{-1}A_{i_1}}\cdots1_{T_{\tau}^{-n}A_{i_n}}dqdp=\nonumber\\
&\left(\int_{\mathbf{R}^2} 1_{A_{i_0}}dqdp\right)\left(\int_{\mathbf{R}^2} 1_{T_{\tau}^{-1}A_{i_1}}dqdp\right)\cdots\left(\int_{\mathbf{R}^2} 1_{T_{\tau}^{-n}A_{i_n}}dqdp\right)
\end{eqnarray}
and since $\int_{\mathbf{R}^2}1_{A}(q,p)dqdp=\mu(A)$ for all $A\in\mathbf{R}^2$ the eq. (\ref{4-3}) implies
\begin{eqnarray}\label{4-4}
\mu(A_{i_0}\cap T_{\tau}^{-1}A_{i_1} \cap \ldots \cap T_{\tau}^{-n}A_{i_n})=\mu(A_{i_0})\mu(T_{\tau}^{-1}A_{i_1})\ldots\mu(T_{\tau}^{-n}A_{i_n})
\end{eqnarray}
Also, since the $T_t$ preserves $\mu$ then one has
\begin{eqnarray}\label{4-5}
\mu(A_{i_0}\cap T_{\tau}^{-1}A_{i_1} \cap ... \cap T_{\tau}^{-n}A_{i_n})=\mu(A_{i_0})\mu(A_{i_1})\ldots\mu(A_{i_n})
\end{eqnarray}
and given that all the elements $A_i$ of $Q_{max}$ have the same volume $\mu(A_i)=\frac{1}{M}=h$ then from (\ref{4-5}) one obtains
\begin{eqnarray}\label{4-6}
\mu(A_{i_0}\cap T_{\tau}^{-1}A_{i_1} \cap \ldots \cap T_{\tau}^{-n}A_{i_n})=h^{n+1}
\end{eqnarray}
Now, from (\ref{4-6}) and the definition (\ref{entropy partition})
one obtains the entropy of $\vee_{j=0}^{n}T_{\tau}^{-j}Q_{max}$, i.e.
\begin{eqnarray}\label{4-7}
&H(\vee_{j=0}^{n}T_{\tau}^{-j}Q_{max})=\\
&-\sum_{(i_0,i_1,\ldots,i_n)}h^{(n+1)}\log h^{(n+1)}=-(n+1)\sum_{(i_0,i_1,\ldots,i_n)}h^{(n+1)}\log h \nonumber
\end{eqnarray}
To complete the calculus one needs to know the number of $(n+1)$--uplas $(i_0,i_1,\ldots,i_n)$. The most simplified situation is to consider that the mixing dynamics is such that for all $n$ and $(i_0,i_1,\ldots,i_n)$ the sets $A_{i_0}\cap T^{-1}A_{i_1} \cap \ldots \cap T^{-n}A_{i_n}$ are all different. In other words, one has $M$ possibilities for $i_0$, the same for $i_1$ and so on. This means that
\begin{eqnarray}\label{4-8}
\sum_{(i_0,i_1,\ldots,i_n)}\leq M^{n+1}
\end{eqnarray}
which expresses that $M^{n+1}$ is an upper bound for the number of $(i_0,i_1,\ldots,i_n)$ that give rise to different subsets
$A_{i_0}\cap T_{\tau}^{-1}A_{i_1} \cap \ldots \cap T_{\tau}^{-n}A_{i_n}$. From the graininess condition (\ref{graininess}) and Eqs. (\ref{4-7})--(\ref{4-8}) one has
\begin{eqnarray}\label{4-9}
&H(\vee_{j=0}^{n}T^{-j}Q_{max})\leq -(n+1)\log h
\end{eqnarray}
This equation states that the entropy of $\vee_{j=0}^{n}T_{\tau}^{-j}Q_{max}$ can grow, at most, as a linear function of the time.
Finally, replacing the supreme in (\ref{KS-entropy2}) by the limit $\lim_{n\rightarrow\infty}\frac{1}{n}H(\vee_{j=0}^{n}T_{\tau}^{-j}Q_{max})$
one obtains
\begin{eqnarray}\label{mixingKS-entropy}
&h_{KS}^{(\tau)}=lim_{n\rightarrow\infty}\frac{1}{n}H(\vee_{j=0}^{n}T_{\tau}^{-j}Q_{max})\leq
lim_{n\rightarrow\infty}\frac{1}{n}\left(-(n+1)\log h\right)=\nonumber\\
&\left(-\log h\right) \left(lim_{n\rightarrow\infty}\frac{n+1}{n}\right)=-\log h
\end{eqnarray}
Now, since $h_{KS}^{(\tau)}=\tau h_{KS}$ then we arrive to our main result of the paper:
\begin{eqnarray}\label{timescale}
h_{KS}\leq -\frac{\log h}{\tau}
\end{eqnarray}
which is the upper bound sought for the KS--entropy in terms of the Planck constant $h$ and the timescale $\tau$.
It should be noted that the equality in (\ref{timescale}) is only satisfied when the dynamics is totally chaotic, which is the case of a large number of chaotic systems such as Sinai billiards \cite{Cas05}, quantum chaotic maps \cite{Ben05}, atoms immersed in a mean electromagnetic field \cite{Cas95}, etc. Therefore, it is necessary to explore the consequences of the mentioned equality. Moreover, when $\Omega$ is not normalized the graininess relation $M h=1$ reads as $M h=\mu(\Omega)$. Then, in the general case one must replace $h$ by $\frac{h}{\mu(\Omega)}=q^{-1}$ with $q$ the quasiclassical parameter. Doing this, the timescale $\tau$ can be expresses as
\begin{eqnarray}\label{timescale1}
\tau = \frac{\log q}{h_{KS}}
\end{eqnarray}
which is nothing but the logarithmic timescale \cite{Cas95}. One final remark that deserves to be mentioned is the following. From (\ref{timescale}) one can see that the upper bound diverges in the classical limit $h\rightarrow0$. This is interpreted by some authors \cite{Cri93,Cri94,Fal03,Ben04,Ben05} as a manifestation of the non commutativity 
where the first order leads to classical chaos and the second one represents a quantum behavior with no chaos at all.

\section{Conclusions}\label{sec conclusions}
We have presented a method for calculating an upper bound of the Kolmogorov--Sinai entropy of a quantum system having a mixing phase space.
The three ingredients that we used were: 1) the natural graininess of the quantum phase space given by the Uncertainty Principle,
2) a time rescaled KS--entropy that allows one to introduce the
characteristic timescale of the system as a parameter, and 3) the factorization property of the mixing correlations given by the lemma \ref{lemma1} .

In summary, our contribution is two--fold. On the one hand, the correspondence between classical and quantum elements of the mixing formalism provides a framework for exporting theorems and results of the classical ergodic theory to quantum language (lemmas \ref{lemma2} and \ref{lemma3}) which is schematized in Fig. \ref{fig:classical-quantum}. On the other hand, the equation (\ref{timescale1}) can be considered as a rigorous proof of the existence of the logarithmic timescale when the dynamics in quantum phase space is fully chaotic, thus providing a theoretical bridge between the ergodic
theory and the graininess of the quantum phase space.

Analogously as was made in \cite{paper0,paper1,paper2,paper3,paper4,paper5}, we hope that the use of more results of the ergodic hierarchy may continue to shed light on the foundations of quantum chaos phenomena in future researches.

\section*{Acknowledgments}
This work was partially supported by CONICET (National Research Council) and Universidad Nacional de La Plata, Argentina.

\appendix
\clearpage 

\section*{Appendices}
\section{Proof of Lemma \ref{lemma1}}
\begin{proof}
First we write $f_{*}$ as a linear combination of characteristic functions, that is $f_{*}=\sum_i \alpha_i1_{C_i}$ with
$C_i\cap C_{i^{\prime}}=\emptyset$ if $i\neq i^{\prime}$ and $\int_{\Gamma}f_{\ast}=\sum_i \alpha_i\mu(C_i)=1$.
Let $A_1$ and $A_2$ be two subsets of the phase space. In particular, we can write
\begin{equation}\label{lemma1-5}
\mu(T_tA\cap B)=C(T_tA,B)+\mu(A)\mu(B)
\end{equation}
where $C(T_tA,B)=\mu(T_tA\cap B)-\mu(A)\mu(B)$.
Hence, on one hand we have
\begin{eqnarray}\label{lemma1-6}
&\sum_i \alpha_i \sum_j \alpha_j\mu(T_tC_i \cap A_1 \cap C_j \cap A_2)=\nonumber\\
&\sum_{i,j} \alpha_i \alpha_jC(T_tC_i,A_1 \cap C_j \cap A_2)+\mu(C_i)\mu(C_j\cap A_1 \cap A_2)\nonumber\\
&=\sum_i \alpha_i \sum_j \alpha_j C(T_tC_i,A_1 \cap C_j \cap A_2)+\nonumber\\
&\sum_i \alpha_i \mu(C_i)\sum_j \alpha_j \mu(C_j\cap A_1 \cap A_2)\\
&=\sum_{i,j} \alpha_i \alpha_j C(T_tC_i,A_1 \cap C_j \cap A_2)+\sum_j \alpha_j \int_{\Gamma}1_{C_j \cap A_1 \cap A_2}\nonumber\\
&=\sum_{i,j}\alpha_i \alpha_j C(T_tC_i,A_1 \cap C_j \cap A_2)+\int_{\Gamma}\sum_j \alpha_j 1_{C_j} 1_{A_1} 1_{A_2}\nonumber\\
&=\sum_{i,j}\alpha_i \alpha_j C(T_tC_i,A_1 \cap C_j \cap A_2)+\int_{\Gamma}f_{\ast} 1_{A_1} 1_{A_2}\nonumber
\end{eqnarray}
On the other hand we also have that
\begin{eqnarray}\label{lemma1-7}
&\sum_i \alpha_i \sum_j \alpha_j\mu(T_tC_i \cap A_1 \cap C_j \cap A_2)=\nonumber\\
&\sum_i \alpha_i \sum_j \alpha_j\mu(T_tC_i \cap T_t(T_{-t}A_1) \cap C_j \cap A_2)=\nonumber\\
&\sum_i \alpha_i \sum_j \alpha_jC(T_t(C_i \cap T_{-t}A_1), C_j \cap A_2)+\nonumber\\
&\sum_i \alpha_i \mu(C_i \cap T_{-t}A_1)\sum_j \alpha_j\mu(C_j\cap A_2)=\nonumber\\
&\sum_i \alpha_i \sum_j \alpha_jC(T_t(C_i \cap T_{-t}A_1), C_j \cap A_2)+\\
&\int_{\Gamma}\sum_i \alpha_i 1_{C_i} 1_{T_{-t}A_1}\int_{\Gamma}\sum_j \alpha_j 1_{C_j}1_{A_2}=\nonumber\\
&\sum_i \alpha_i \sum_j \alpha_jC(T_t(C_i \cap T_{-t}A_1), C_j \cap A_2)+\nonumber\\
&\int_{\Gamma}f_{\ast} 1_{T_{-t}A_1}\int_{\Gamma}f_{\ast}1_{A_2}=\nonumber\\
&\sum_i \alpha_i \sum_j \alpha_jC(T_t(C_i \cap T_{-t}A_1), C_j \cap A_2)+\nonumber\\
&\int_{T_{-t}A_1}f_{\ast}\int_{\Gamma}f_{\ast}1_{A_2}\nonumber
\end{eqnarray}
Now by the property of the Frobenius-Perron operator $P_t$ and since $f_{\ast}$ is a fixed point of $P_t$ (i.e. $P_tf_{\ast}=f_{\ast}$) we have
\begin{equation}\label{lemma1-8}
\int_{T_{-t}A_1}f_{\ast}=\int_{A_1}P_tf_{\ast}=\int_{A_1}f_{\ast}=\int_{\Gamma}f_{\ast}1_{A_1}
\end{equation}
Then using (\ref{lemma1-8}) we can recast (\ref{lemma1-7}) as
\begin{eqnarray}\label{lemma1-9}
&\sum_i \alpha_i \sum_j \alpha_j\mu(T_tC_i \cap A_1 \cap C_j \cap A_2)=\nonumber\\
&\sum_i \alpha_i \sum_j \alpha_jC(T_t(C_i \cap T_{-t}A_1), C_j \cap A_2)+\\
&\int_{\Gamma}f_{\ast}1_{A_1}\int_{\Gamma}f_{\ast}1_{A_2}\nonumber
\end{eqnarray}
Now due the mixing correlation of eq. (\ref{mixing1}) we can take the limit $t\rightarrow\infty$ in eqns. (\ref{lemma1-6}) and (\ref{lemma1-9}) and we obtain that
$C(T_tC_i,A_1 \cap C_j \cap A_2)$ and $C(T_t(C_i \cap T_{-t}A_1)$ tend to zero. Therefore, we have
\begin{eqnarray}\label{lemma1-10}
&\lim_{t\rightarrow\infty}\sum_i \alpha_i \sum_j \alpha_j\mu(T_tC_i \cap A_1 \cap C_j \cap A_2)=\\
&\int_{\Gamma}f_{*}1_{A_1}1_{A_2}=\int_{\Gamma}f_{*}1_{A_1}\int_{\Gamma}f_{*}1_{A_2}\nonumber
\end{eqnarray}
If we have $n$ characteristic functions $1_{A_1},1_{A_2},...,1_{A_n}$ we simply apply $n-1$ times the Eq. (\ref{lemma1-10}) to obtain
\begin{eqnarray}\label{lemma1-11}
&\int_\Gamma f_{*} 1_{A_1}1_{A_2}...1_{A_n}=\int_\Gamma f_{*} 1_{A_1}1_{A_2\cap...\cap A_n}=\nonumber\\
&\int_\Gamma f_{*} 1_{A_1}\int_\Gamma f_{*} 1_{A_2\cap...\cap A_n}=\\
&\int_\Gamma f_{*} 1_{A_1}\int_\Gamma f_{*} 1_{A_2}\int_\Gamma f_{*} 1_{A_3\cap...\cap A_n}
=...^{n-3 \ times}...\nonumber\\
&=\int_\Gamma f_{*} 1_{A_1}\int_\Gamma f_{*} 1_{A_2}...\int_\Gamma f_{*} 1_{A_n}\nonumber
\end{eqnarray}
\end{proof}

\section{Proof of Lemma \ref{lemma2}}
\begin{proof}
Let $s\in\mathbf{R}$ be a real number. Then replacing $\hat{O}$ by $\hat{U}_{s}^{\dagger}\hat{O}\hat{U}_{s}$ in eq. (\ref{qmixing1}) we have
\begin{eqnarray}\label{qmixing2}
\lim_{t\rightarrow\infty}\langle\hat{U}_{s}^{\dagger}\hat{O}\hat{U}_{s}\rangle_{\hat{\rho}(t)}-\langle\hat{U}_{s}^{\dagger}\hat{O}\hat{U}_{s}\rangle_{\hat{\rho}_{*}}=0
\end{eqnarray}
Now applying trace properties we can rewrite eq. (\ref{qmixing2}) as
\begin{eqnarray}\label{qmixing3}
\lim_{t\rightarrow\infty}\langle\hat{O}\rangle_{\hat{\rho}(t+s)}-\langle\hat{O}\rangle_{\hat{U}_{s}\hat{\rho}_{*}\hat{U}_{s}^{\dagger}}=0
\end{eqnarray}
where
\begin{eqnarray}\label{qmixing4}
\lim_{t\rightarrow\infty}\langle\hat{O}\rangle_{\hat{\rho}(t+s)}=\lim_{t\rightarrow\infty}\langle\hat{O}\rangle_{\hat{\rho}(t)}=
\langle\hat{O}\rangle_{\hat{\rho}_{*}}
\end{eqnarray}
Now from (\ref{qmixing3}) and (\ref{qmixing4}) it follows that
$\langle\hat{O}\rangle_{\hat{U}_{s}\hat{\rho}_{*}\hat{U}_{s}^{\dagger}}=\langle\hat{O}\rangle_{\hat{\rho}_{*}}$ for all observable $\hat{O}$, which means that
\begin{eqnarray}\label{lemma1-17}
\hat{U}_s\hat{\rho}_{\ast}\hat{U}_s^{\dag}=\hat{\rho}_{\ast} \ \ \ \forall s\in\mathbf{R}
\end{eqnarray}
\end{proof}

\section{Proof of Lemma \ref{lemma3}}
\begin{proof}
By applying the definition of Frobenius-Perron operator (i.e. $\int_A P_tf=\int_{T_{-t}A}f$) to the Wigner function $W_{\hat{\rho}_{\ast}}(q,p)$, using the lemma \ref{lemma2} and the Wigner property (\ref{wigner8}) we have
\begin{eqnarray}\label{lemma1-18}
&\int_{A}P_t W_{\hat{\rho}_{\ast}}(q,p)dqdp=\int_{T_{-t}A}W_{\hat{\rho}_{\ast}}(q,p)dqdp=\nonumber\\
&\int_{\mathbf{R}}W_{\hat{\rho}_{\ast}}(q,p)1_{T_{-t}A}(q,p)dqdp=\nonumber\\
&\int_{\mathbf{R}}W_{\hat{\rho}_{\ast}}(q,p)\widetilde{W}_{\hat{U}_t^{\dag}\hat{I}_{A}\hat{U}_t}(q,p)dqdp=\nonumber\\
&Tr(\hat{\rho}_{\ast}\hat{U}_t^{\dag}\hat{I}_{A}\hat{U}_t)=Tr(\hat{U}_t\hat{\rho}_{\ast}\hat{U}_t^{\dag}\hat{I}_{A})=
Tr(\hat{\rho}_{\ast}\hat{I}_{A})=\\
&\int_{\mathbf{R}}W_{\hat{\rho}_{\ast}}(q,p)\widetilde{W}_{{I}_{A}}(q,p)dqdp=\nonumber\\
&\int_{\mathbf{R}}W_{\hat{\rho}_{\ast}}(q,p)1_A(q,p)dqdp=\int_{A}W_{\hat{\rho}_{\ast}}(q,p)dqdp\nonumber
\end{eqnarray}
where we have also used that $\widetilde{W}_{\hat{U}_t^{\dag}\hat{I}_{A}\hat{U}_t}(q,p)=1_{T_{-t}A}(q,p)$ being
$\hat{I}_{A}$ the operator whose Weyl symbol is the characteristic function $1_A(q,p)$, i.e. $\widetilde{W}_{\hat{I}_{A}}(q,p)=1_{A}(q,p)$.
Then from the eq. \ref{lemma1-18} it follows that
$P_t W_{\hat{\rho}_{\ast}}(q,p)=W_{\hat{\rho}_{\ast}}(q,p)$.
\end{proof}
\end{document}